\documentclass[11pt,a4paper]{article}
\usepackage{jheppub}
\usepackage[T1]{fontenc}

\allowdisplaybreaks

\newcommand{\0}{SO(10)}

\newcommand{\be}{\begin{equation}}
\newcommand{\ee}{\end{equation}}
\newcommand{\bea}{\begin{eqnarray}}
\newcommand{\eea}{\end{eqnarray}}
\newcommand{\ba}{\begin{array}}
\newcommand{\ea}{\end{array}}

\title{On the Seesaw Scale in Supersymmetric  SO(10) Models}
\author{Zhenxin Ren}
\author{and Da-Xin Zhang}
\affiliation{School of Physics and State Key Laboratory of Nuclear Physics and Technology, \\
Peking University, Beijing 100871, China}
\emailAdd{1301210106@pku.edu.cn}
\emailAdd{dxzhang@pku.edu.cn}
\abstract{
The seesaw mechanism, which is responsible for the description of neutrino masses and mixing,
requires a scale lower than the unification scale.
We propose a new model with spinor superfields playing important roles to generate this seesaw scale,
with special attention paid on the Goldstone mode of the $U(1)_{B-L}$ symmetry breaking.
}
\begin{document}
\maketitle
\flushbottom
\pagenumbering{arabic}

\section{Introduction}

Supersymmetric (SUSY) Grand Unified Theories (GUTs) of SO(10)\cite{so10a,so10b} are very important
in searching for the new physics beyond the Standard Model (SM).
First, the running behaviors of the three gauge couplings suggest that
in the Minimal SUSY SM (MSSM) they unify at a scale around $2\times 10^{16}$GeV\cite{unif1,unif2,MSSM1,MSSM2,MSSM3,MSSM4}
which is called the GUT scale $\Lambda_{GUT}$.
Second,  every generation of   quarks and leptons
 are contained in a spinor
representation $\textbf{16}$ of  SO(10) which also has the component of a right-handed neutrino.
Since in SO(10) the extra $U(1)_{B-L}$ symmetry needs to be broken with the corresponding
Vacuum Expectation Value (VEV)
giving the Majorana masses to the right-handed neutrinos,
the low energy data on neutrino masses and mixing can be beautifully described by
the seesaw mechanism\cite{seesawi1,seesawi2,seesawi3,seesawi4,seesawi5,seesawii1,seesawii2,seesawii3,seesawii4}.

Detailed studies\cite{sato,aulakh4554,msso10a,msso10b,bajc,msso10d,malinsky,aulakh120,bert}
suggest that there are difficult problems to be solved in the SUSY SO(10) models.
First, as in other SUSY GUT models, proton decays are mainly induced by the dimension-five
operators mediated by the color-triplet Higgsinos whose masses of the order $\Lambda_{GUT}$
are not large enough to suppress proton decays\cite{su5pd1,su5pd2,su5pd3,su5pd4}.
Second, in SO(10) there is a conflict between the seesaw sale and gauge coupling unification.
The low energy neutrino data suggest that the right-handed neutrinos couple to field with
a VEV of the order $\Lambda_{seesaw}\sim 10^{14}$GeV,
while gauge coupling unification  disfavors any intermediate scale\cite{bajc}.
Third, in the MSSM there is a pair of weak doublets whose masses are negligibly smaller
compared to other masses  in the same representations
which
are of the order $\Lambda_{GUT}$. The large Doublet-Triplet Splitting (DTS) cannot be
simply realized without introducing delicate mechanisms.

In recent studies, by going beyond the minimal model,
all these major difficulties have been solved in a somehow complicated renormalizable model\cite{cz2}.
Usually in renormalizable models the $U(1)_{B-L}$ symmetry is  broken by the tensor representations
\textbf{126} $+\overline{\textbf{126}}$.
In \cite{cz2} there are two pairs of Higgs in  \textbf{126} $+\overline{\textbf{126}}$
responsible to the $U(1)_{B-L}$ breaking following \cite{dlz2}.
Only one $\overline{\textbf{126}}$ with a $\Lambda_{seesaw}$ VEV
couples to the MSSM matter superfields,
while the other VEVs responsible for the $U(1)_{B-L}$ breaking
take $\Lambda_{GUT}$ values and SO(10) breaks directly into the SM gauge symmetry.
Since $\Lambda_{seesaw}$
is not a real intermediate scale, the gauge coupling unification maintains.
Furthermore, although all the color triplets in this model have $\Lambda_{GUT}$ masses,
their effective triplet masses are found to be $\frac{\Lambda_{GUT}^2}{\Lambda_{seesaw}}$
which are used to suppress proton decays.
Also, natural DTS is realized through the Dimopoulos-Wilczek (DW) Mechanism
\cite{dw1,dw2,dw3,lee,cz} of missing VEV.
The model \cite{cz2} has been examined numerically to be realistic in \cite{czy}.

It can be seen that $\Lambda_{seesaw}$ plays a very important role
in realistic models.
In \cite{cz2} it simply uses a
global symmetry whose breaking scale is taken around $\Lambda_{seesaw}$.
The  seesaw mechanism of type-I explains the typical low energy neutrino masses as
\be
m_\nu=-\frac{v_{EW}^2}{\Lambda_{seesaw}},\label{ss}
\ee
where $v_{EW}$ is the electroweak scale of the SM at which the Dirac masses for the neutrinos are taken.
However, if we rewrite (\ref{ss}) as
\be
\frac{m_\nu}{v_{EW}}=-\frac{v_{EW}}{\Lambda_{seesaw}},\label{ss2}
\ee
it only means the replacement of a very small number  $10^{-10}$ on the LHS  by another one on the RHS.
To be more realistic, we need to give an explanation on this value $\Lambda_{seesaw}$.
This has been done in \cite{lz} where a third pair of \textbf{126} $+\overline{\textbf{126}}$
are introduced which couple to a SO(10) singlet charged under an
anomalous $U(1)_A$ whose VEV is naturally at the string or the reduced-Planck scale
$\Lambda_{st}\sim 10^{18}$GeV according to the Green-Schwarz mechanism\cite{green1,green2,green3,green4}.
Consequently,
the seesaw scale is generated as $\Lambda_{seesaw}\sim \frac{\Lambda_{GUT}^2}{\Lambda_{st}}$.

The  seesaw scale is so important that we need to know if there are other models
which can generate  it, especially through the Green-Schwarz mechanism.
Usually in the renormalizable SO(10) models, only the tensors  \textbf{126} $+\overline{\textbf{126}}$
are responsible for the $U(1)_{B-L}$ breaking while the spinors
\textbf{16}$+\overline{\textbf{16}}$ cannot appear.
However, this is not necessarily true if these spinor Higgs differ explicitly from the
three generations of the matter superfields.

In the present work we propose a mechanism with the \textbf{16}$+\overline{\textbf{16}}$
helping to generate the seesaw scale through the Green-Schwarz mechanism.
Instead of using three pairs of  \textbf{126} $+\overline{\textbf{126}}$
as in \cite{lz},
we find that the minimal setting needs two pairs of them and two pairs of \textbf{16}$+\overline{\textbf{16}}$.
In the meantime, the successful mechanism of suppressing proton decay
is not lost. The DTS problem, however, will not be naturally realized which is beyond the present study.

In Section 2, we will analyze the simple SUSY SO(10) models with one pair of
\textbf{126}$+\overline{\textbf{126}}$ or
\textbf{16}$+\overline{\textbf{16}}$ to break $U(1)_{B-L}$.
In Section 3, we will
extend the minimal models to incorporate the mechanisms of the proton decay suppression
and the seesaw scale generation.
In Section 4 we will analyze the simplest model with both \textbf{126}$+\overline{\textbf{126}}$
and \textbf{16}$+\overline{\textbf{16}}$ existing together.
We give the general form of the Goldstone mode for the $U(1)_{B-L}$ symmetry breaking whose proof is
given  in the Appendix.
In Section 5 we will propose a new model of generating the seesaw scale. In Section 6 we will summarize.

\section{The simple SUSY SO(10) Models}

We start with analyzing the $U(1)_{B-L}$ breaking in the simple SO(10) models.
The minimal SUSY SO(10) contains Higgs superfields in $H(\textbf{10}),\Delta(\textbf{126}) +\overline{\Delta}(\overline{\textbf{126}}), \Phi(\textbf{210})$\cite{msso10a,msso10b,bajc,msso10d}.
Here, $H(\textbf{10})$ (and also $D(\textbf{120})$ in some extended models\cite{malinsky,aulakh120})
is not responsible for
the GUT symmetry breaking and is irrelevant.
The $\Phi(\textbf{210})$ can be replaced by $A(\textbf{45})$ +$E(\textbf{54})$
as an alternative\cite{aulakh4554}, and we will use the later as examples.
The general superpotential relevant for the $U(1)_{B-L}$ breaking is
\be
\overline{\Delta} (m_{\Delta}+\lambda A)  \Delta.
\ee
Labeling the VEVs using their representations under the $SU(4)_C\times SU(2)_L\times SU(2)_R$
subgroup of SO(10), they are
\be
\overline{v_R}\equiv \langle\overline{\Delta}{(10,1,3)}\rangle,
~v_R\equiv\langle\Delta{(\overline{10},1,3)}\rangle,
~A_1\equiv \langle A{(1,1,3)}\rangle,
~A_2\equiv \langle A{(15,1,1)}\rangle.
\ee
Preserving SUSY at high energy requires the D-flatness condition
\be
\overline{v_R}=v_R
\ee
and the F-flatness conditions
\be
0=F_{\overline{v_R}}=M_\Delta v_R, ~~0=F_{v_R}=\overline{v_R}M_\Delta,
\ee
where $M_\Delta=m_{\Delta}+\lambda A_0$ and $A_0=-\frac{1}{5} A_1-\frac{\sqrt{3}}{5\sqrt{2}}A_2$.
For nonzero $\overline{v_R}=v_R$,
\be
M_\Delta=0\label{MDelta}
\ee
which determines $A_0$.
Full determinations of  all the VEVs need to use the complete superpotential which is simple and is
irrelevant here.

The singlets of the SM gauge group must contain a massless eigenstate which is
the Goldstone mode for the $U(1)_{B-L}$ breaking.
This mass matrix for the SM singlets is
\be
\left[ \ba{cc|cc}
0& M_\Delta& -\lambda\frac{1}{5} \overline{v_R}  & ~-\lambda\frac{\sqrt{3}}{5\sqrt{2}} \overline{v_R}\\
M_\Delta&0 & -\lambda\frac{1}{5}v_R & ~-\lambda\frac{\sqrt{3}}{5\sqrt{2}}v_R\\
\hline
-\lambda\frac{1}{5} \overline{v_R} & ~-\lambda\frac{1}{5}v_R~ &\star & \star\\
-\lambda\frac{\sqrt{3}}{5\sqrt{2}} \overline{v_R} & ~-\lambda\frac{\sqrt{3}}{5\sqrt{2}}v_R~ & \star&\star\\
 \ea\right],\label{star}
\ee
where both the  columns and the rows are ordering as $\Delta,
\overline{\Delta}, A_1, A_2$.
The ``$\star$''s stand for irrelevant quantities.
The massless eigenstate is easy to find to be a combination of SM singlets in $\Delta$ and $\overline{\Delta}$,
since the upper two rows are not independent, neither are the left two columns,
when (\ref{MDelta}) applies.
In this simple situation, the existence of the Goldstone mode can be taken as an
automatic result of the F-flatness conditions.
In a full model $E(\textbf{54})$ is also needed for the GUT symmetry breaking, but including $E$ in
the mass matrix  for the SM singlets will not
change the above results since it does not couple with the SM singlets in $\Delta,
\overline{\Delta}$.
$E$ and the ``$\star$''s in (\ref{star}) do not enter into the eigenvalue equation of the Goldstone mode.

Now we use $\Psi(\textbf{16})+\overline{\Psi}(\overline{\textbf{16}})$
to break $U(1)_{B-L}$. The  VEVs are denoted as
\be
\psi=\langle \Psi(\overline{4},1,2)\rangle, ~~\overline{\psi}=\langle \overline{\Psi}(4,1,2)\rangle,
\ee
and the relevant superpotential  is
\be
\overline{\Psi} (m_{\Psi}+\eta A)  \Psi.
\ee
The D-flatness condition is
\be
\overline{\psi}=\psi
\ee
and the F-flatness conditions are
\be
0=F_{\overline{\psi}}=M_\Psi \psi, ~0=F_{\psi}=\overline{\psi}M_\Psi,
\ee
where $M_\Psi=m_{\Psi}+\eta A_0^\prime$ and $A_0^\prime= -2 A_1-\sqrt{6}A_2$. For nonzero
$\overline{\psi}=\psi$,
\be
M_\Psi=0
\ee
determining $A_0^\prime$. The mass matrix for the SM singlets is
\be
\left[ \ba{cc|cc}
0 & M_\Psi& -\eta 2 \overline{\psi}  & ~-\eta\sqrt{6}\overline{\psi}~\\
M_\Psi & 0& -\eta 2\psi & ~-\eta\sqrt{6}\psi~ \\
\hline
-\eta 2 \overline{\psi} & ~-\eta 2\psi~ &\star & \star\\
-\eta\sqrt{6} \overline{\psi} & ~-\eta\sqrt{6}\psi~& \star&\star\\
 \ea\right],
\ee
where both the columns and the rows are ordering as
$\Psi, \overline{\Psi}, A_1, A_2$.
Again, the Goldstone mode exists following the F-flatness conditions.
Also, we have not included explicitly $E$ which does not couple with the SM singlets in $\Psi, \overline{\Psi}$
and hence will not change the above results.

\section{Generation of the seesaw scale in the realistic models}

The simple models  in  Section 2 cannot be realistic. To suppress proton decay in the renormalizable models,
the Higgs superfields which couple with the matter superfields in the MSSM
need to be extended.
The Yukawa sector is denoted as
\be
W_{Yukawa}=\sum_{i,j=1,2,3}\Psi_i\Psi_j
\left( Y_{10,ij} H_1(10)+Y_{120,ij} D_1(120)+Y_{126,ij} \overline{\Delta}_1(\overline{126})\right),\label{yuka}
\ee
where $\Psi_{1,2,3}$ are the MSSM matter superfields which do not contribute to the GUT symmetry breaking.
Denoting those Higgs superfields which do not contribute
to the Yukawa couplings  as $H_2(10), D_2(120), \overline{\Delta}_2(\overline{126}), {\Delta}_2({126})$,
the mass matrix for the color-triplets is divided into $6\times 6$ sub-matrices,
\be
\left[ \ba{cc}
0& \hat{M}^T_{12}\\
\hat{M}^T_{21}&\hat{M}^T_{22}
\ea\right]\label{3.2}
\ee
following the Higgs superpotential
\bea
&&m_{H12}H_1H_2+m_{D12}D_1D_2+m_{\Delta 12}\overline{\Delta}_1{\Delta}_2+m_{\Delta 21}\overline{\Delta}_2 {\Delta}_1
\nonumber\\
&+&A\left( H_1H_2+H_1D_2+H_2D_1+D_1D_2
+D_1\overline{\Delta}_2+D_1{\Delta}_2+D_2\overline{\Delta}_1+D_2{\Delta}_1
+\overline{\Delta}_1{\Delta}_2+\overline{\Delta}_2{\Delta}_1\right)
\nonumber\\
&+&X \left( H_2^2+D_2^2+\overline{\Delta}_2{\Delta}_2 \right).\label{dlz2}
\eea
Here we have suppressed all dimensionless couplings for concise.
Also, possible couplings
\be
EH_i H_j, ~ED_i D_j, ~E\Delta_i\Delta_j, ~E\overline{\Delta}_i\overline{\Delta}_j
\ee
need to be included if allowed.

For the SO(10) singlet $X$ given a VEV $\ll  \Lambda_{GUT}$ and all the other VEVs and mass parameters are  at the
GUT scale,
all the triplets have GUT scale masses.
In the mass matrix,
\be
\hat{M}^T_{12}\sim \Lambda_{GUT}, ~\hat{M}^T_{21}\sim \Lambda_{GUT},  ~\hat{M}^T_{22}\sim X.
\ee
The effective triplet mass matrix, which is got
by integrating out those fields which do not appear in the Yukawa superpotential,
has all entries $\frac{\Lambda_{GUT}^2}{X}$,
thus the amplitudes for proton decay mediated by the
color-triplet Higgsinos are suppressed for $X \ll  \Lambda_{GUT}$.

The superpotential (\ref{dlz2}) must be protected by extra symmetries, otherwise unwanted
terms reappear and no suppression of proton decay can be assured.
Charged under a global symmetry, in \cite{dlz2} the VEV of $X$ is taken to be $10^{-2}\Lambda_{GUT}$.
Consequently,
the D-flatness condition is
\be
|\overline{v_{1R}}|^2+|\overline{v_{2R}}|^2=|v_{1R}|^2+|v_{2R}|^2,
\ee
and F-flatness conditions are
\be
0=\left[\overline{v_{1R}}, ~\overline{v_{2R}}\right]
\left[\ba{cc} 0&M_{\Delta 12}\\M_{\Delta 21}& X\ea \right] ,
~0=\left[\ba{cc} 0&M_{\Delta 12}\\M_{\Delta 21}& X\ea \right]
\left[\ba{c} v_{1R}\\v_{2R}\ea\right],\label{2X2}
\ee
where $M_{\Delta,ij}=m_{\Delta,ij}+\lambda_{ij} A_0$.

The key point is that the $2\times 2$ matrix in (\ref{2X2}) has one but only one zero eigenvalue.
Then, in the symmetric mass matrix of the SM singlets,
ordering as $\Delta_{1}, \Delta_{2},
\overline{\Delta}_1, \overline{\Delta}_2, A_1, A_2$,
\be
\left[
\ba{cccc|cc}
~&~&0&M_{\Delta 21}& -\lambda_{21}\frac{1}{5} \overline{v_{2R}}
& ~-\lambda_{21}\frac{\sqrt{3}}{5\sqrt{2}} \overline{v_{2R}}~\\
~&~&M_{\Delta 12}&X & -\lambda_{12}\frac{1}{5} \overline{v_{1R}}
& ~-\lambda_{12}\frac{\sqrt{3}}{5\sqrt{2}} \overline{v_{1R}}~\\
0&M_{\Delta 12}&~&~& -\lambda_{12}\frac{1}{5}v_{2R}
& ~-\lambda_{12}\frac{\sqrt{3}}{5\sqrt{2}}v_{2R}~\\
M_{\Delta 21}&X &~&~& -\lambda_{21}\frac{1}{5}v_{1R}
& ~-\lambda_{21}\frac{\sqrt{3}}{5\sqrt{2}}v_{1R}~\\
\hline
-\lambda_{21}\frac{1}{5} \overline{v_{2R}} &~-\lambda_{12}\overline{v_{1R}}
&-\lambda_{12}\frac{1}{5} v_{2R} &-\lambda_{21}\frac{1}{5}v_{1R}&\star&\star\\
~-\lambda_{21}\frac{\sqrt{3}}{5\sqrt{2}}\overline{v_{2R}} & ~-\lambda_{12}\frac{\sqrt{3}}{5\sqrt{2}}\overline{v_{1R}}
& ~-\lambda_{12}\frac{\sqrt{3}}{5\sqrt{2}} v_{2R}& ~-\lambda_{21}\frac{\sqrt{3}}{5\sqrt{2}}v_{1R}&\star&\star
\ea \right].
\ee
The upper-left $4\times 4$ sub-matrix has two zero eigenvalues following (\ref{2X2}),
while in the upper-right $4\times 2$ sub-matrix there is only one independent row
and in the lower-left $2\times 4$ sub-matrix there is only one independent column.
Consequently, there is only one massless eigenstate which is the Goldstone mode whose
components are all from the $B-L$ charged fields $\Delta_{1,2}$ and $\overline{\Delta}_{1,2}$.

Note that $M_{\Delta 12}$ and $M_{\Delta 21}$ cannot be zero simultaneously without fine-tuning parameters.
We choose  $M_{\Delta 21}=0$ and solve the D- and F-flatness conditions, then
\be
\overline{v_{1R}}=-\frac{X}{M_{\Delta 12}}\overline{v_{2R}},
~~\overline{v_{2R}}\sim v_{1R}\sim \Lambda_{GUT}, ~~v_{2R}=0.
\ee
It is $\overline{v_{1R}}$ which gives the masses to the right-handed neutrinos,
and this seesaw scale is now generated through the VEV of the SO(10) singlet $X$.

In \cite{dlz2}, although the link between the seesaw scale and the proton decay suppression has been  setup,
the VEV of the SO(10) singlet is simply put in by hand and is not a satisfactory.
The global symmetry has been further replaced by an anomalous $U(1)_A$ symmetry in \cite{lz}, where
a third pair of $\overline{\Delta}_3+{\Delta}_3$ are introduced
so that the last term (the term containing $X$) in (\ref{dlz2}) is replaced by
\be
A^\prime( H_2+D_2)  ( \overline{\Delta}_3+{\Delta}_3 )+ A^\prime\overline{\Delta}_2\Delta_3+A^\prime\overline{\Delta}_3{\Delta}_2
+S \overline{\Delta}_3{\Delta}_3,
\ee
where $A^\prime$ is a new \textbf{45} and $S$ is a SO(10) singlet.
Due to the Green-Schwarz mechanism, the VEV of the $S$ is taken to be the string  scale
$\Lambda_{st}\sim 10\Lambda_{GUT}$\cite{green11,green12}.
The D-flatness condition is
\be
|\overline{v_{1R}}|^2+|\overline{v_{2R}}|^2+|\overline{v_{3R}}|^2=|v_{1R}|^2+|v_{2R}|^2+|v_{3R}|^2,
\ee
and the F-flatness conditions are
\be
0=\left[\overline{v_{1R}}, ~\overline{v_{2R}}, ~\overline{v_{3R}}\right]
\left[\ba{ccc} 0&M_{\Delta 12}&0\\M_{\Delta 21}&0& M_{\Delta 23}\\0&M_{\Delta 32}&S\ea\right],
~0=\left[\ba{ccc} 0&M_{\Delta 12}&0\\M_{\Delta 21}&0& M_{\Delta 23}\\0&M_{\Delta 32}&S\ea\right]
\left[\ba{c} v_{1R}\\v_{2R}\\v_{3R}\ea\right],\label{lzmatrix}
\ee
where $M_{\Delta 23}\sim A_0^\prime, ~M_{\Delta 32} \sim A_0^\prime$.
Then, the $U(1)_{B-L}$ symmetry breaking requires
the determinant of the $3\times 3$ matrix in (\ref{lzmatrix}) is zero so that it has one massless eigenstate.
Again we choose $M_{\Delta 21}=0$ which gives
\be
\overline{v_{1R}}=\frac{M_{\Delta 23}M_{\Delta 32}}{SM_{12}}\overline{v_{2R}},
~\overline{v_{3R}}=-\frac{M_{\Delta 23}}{S}\overline{v_{2R}},
~\overline{v_{2R}}\sim v_{1R}\sim \Lambda_{GUT}, ~v_{2R}=v_{3R}=0.
\ee
Solving all the other  F-flatness conditions shows that $A_0^\prime\sim \frac{1}{\sqrt{10}} \Lambda_{GUT}$,
then the seesaw scale $\overline{v_{1R}}\sim 10^{-2} \Lambda_{GUT}$ is generated.
The mechanism of proton decay suppression keeps working, since in getting the effective triplet mass matrix,
a first step of integrating out $\overline{\Delta}_3+{\Delta}_3$ is needed, which amounts to
replacing $X$ in the simple model by $\frac{A^{\prime 2}}{\Lambda_{st}}\sim 10^{-2}\Lambda_{GUT}$.

In the models using only $\Psi(\textbf{16})+\overline{\Psi}(\overline{\textbf{16}})$s
to break the $U(1)_{B-L}$, the seesaw scale can be generated similarly.
However, such kind of models are non-renormalizable.
In the next Section we will study in a renormalizable model where both \textbf{126}$+\overline{\textbf{126}}$
and $\Psi(\textbf{16})+\overline{\Psi}(\overline{\textbf{16}})$ are present.

\section{Breaking $U(1)_{B-L}$ in models with both \textbf{126}$+\overline{\textbf{126}}$ and
\textbf{16}$+\overline{\textbf{16}}$}

Now we  include both a pair of  \textbf{126}$+\overline{\textbf{126}}$
and a pair of  $\Psi(\textbf{16})+\overline{\Psi}(\overline{\textbf{16}})$
in a same model.
We will have some general observation on the $U(1)_{B-L}$ breaking in this case.

The relevant superpotential is
\be
\overline{\Delta}M_{\Delta}\Delta+\overline{\Psi} M_{\Psi}\Psi+\overline{\Delta}\Psi\Psi+\Delta\overline{\Psi}\overline{\Psi}.
\ee
The D-flatness condition is
\be
2|\overline{v_R}|^2+ |\overline{\psi}|^2=2|{v_R}|^2+ |{\psi}|^2,
\ee
and the F-flatness conditions are
\bea
0=F_{\overline{v_R}}&=&M_{\Delta}v_R+\psi\psi, \nonumber\\
0=F_{v_R}&=&\overline{v_R} M_{\Delta}+\overline{\psi}\overline{\psi}, \nonumber\\
0=F_{\overline{\psi}}&=&M_{\Psi}\psi+2v_R\overline{\psi}, \nonumber\\
0=F_{\psi}&=&\overline{\psi} M_{\Psi}+2\overline{v_R}\psi.      \label{F10126}
\eea
Unlike in the models discussed in the previous Sections,
the equations in (\ref{F10126}) are nonlinear and no simple solutions can be directly read off.

Ordering the bases as $\Delta,\overline{\Delta},\Psi,\overline{\Psi}, A_1, A_2$,
the symmetric mass matrix for the SM singlets is
\be
\left[\ba{cccc|cc}
0 & M_\Delta & 0 & 2\overline{\psi}
&-\lambda\frac{1}{5} \overline{v_R}  & ~-\lambda\frac{\sqrt{3}}{5\sqrt{2}} \overline{v_R}\\
M_\Delta & 0 & 2\psi &0
&-\lambda\frac{1}{5}v_R & ~-\lambda\frac{\sqrt{3}}{5\sqrt{2}}v_R\\
0 & 2\psi & 2\overline{v_R} & M_\Psi
&-\eta 2 \overline{\psi}  & ~-\eta\sqrt{6}\overline{\psi}~\\
2 \overline{\psi} & 0 & M_\Psi & 2v_R
& -\eta 2\psi & ~-\eta\sqrt{6}\psi~ \\
\hline
-\lambda\frac{1}{5} \overline{v_R} & ~-\lambda\frac{1}{5}v_R &
-\eta 2 \overline{\psi} & ~-\eta 2\psi~ &\star & \star\\
-\lambda\frac{\sqrt{3}}{5\sqrt{2}} \overline{v_R} & ~-\lambda\frac{\sqrt{3}}{5\sqrt{2}}v_R~ &
-\eta\sqrt{6} \overline{\psi} & ~-\eta\sqrt{6}\psi~& \star&\star\\
\ea\right].\label{16126}
\ee
Following(\ref{F10126}), the upper four rows in (\ref{16126})
can be combined into a row with all its entries being zeros as the Goldstone mode.
Explicitly,
the Goldstone mode is
\be
\frac{2v_R}{N} \Delta-\frac{2\overline{v_R}}{N} \overline{\Delta}
+\frac{\psi}{N} \Psi-\frac{\overline{\psi}}{N}\overline{\Psi}\label{goldstone}
\ee
whose physical meaning is very obvious.
The factors ``2'' and/or ``-'' correspond to the $U(1)_{B-L}$ charges.
This simply follows the F-flatness conditions (\ref{16126}).
Here
\be
N=\left(|2v_R|^2 +|2\overline{v_R}|^2 +|\psi|^2 +|\overline{\psi}|^2\right)^{1/2}\nonumber
\ee
is the normalization factor.
The Goldstone modes in the simple models of Section 2 are special cases of (\ref{goldstone}).
Also, it can be found that in the models discussed in Section 3,
those SM singlets with zero VEVs ($v_{2R}=0$ or $v_{2R}=v_{3R}=0$) do not enter into
the Goldstone constituents. We will give a simple proof for the formula of the Goldstone's constituents
in the Appendix.

\section{The present model}\label{general}

In constructing models which successfully suppress proton decay and generate the seesaw scale
through the Green-Scwarz mechanism,
the mass matrix for the color-triplets needs to be the form of (\ref{3.2}) so that
two pairs of \textbf{126}/$\overline{\textbf{126}}$ are needed,
and the sub-matrix $\hat{M}^T_{22}$ need to be generated through the couplings of
$\Delta_{1,2}/\overline{\Delta}_{1,2}$ with several pairs of $\Psi/\overline{\Psi}$.

The superpotential for the $U(1)_{B-L}$ breaking is
\begin{equation}\label{superp}
\overline{\Delta}_1 M_{\Delta 12}\Delta_2+\overline{\Delta}_2 M_{\Delta 21}\Delta_1
+\overline{\Psi}_k M_{\Psi_{kl}}\Psi_l
+f_{lk}^i\overline{\Delta}_i\Psi_l\Psi_k+g_{lk}^j\Delta_j\overline{\Psi}_k\overline{\Psi}_l ,
\end{equation}
where $f$'s and $g$'s are dimensionless couplings.
$M_{\Delta 12}$ and $M_{\Delta 21}$ need to be taken at the GUT scale for
the particles in $\Delta_{1,2}/\overline{\Delta}_{1,2}$ to have masses of this scale
so that  we can take $M_{\Delta ij}=m_{\Delta ij}+\lambda_{ij} A_0$,
while $M_{\Psi_{kl}}$'s must involve couplings with a large VEV of SO(10) singlet and their explicit forms have
not been determined yet.

The D-flatness condition is
\be
2|\overline{v_{1R}}|^2+2|\overline{v_{2R}}|^2+\sum_k |\overline{\psi_k}|^2
=2|v_{1R}|^2+2|v_{2R}|^2+\sum_k |\psi_k|^2,\label{Dflat}
\ee
while the F-flatness conditions are
\begin{eqnarray}
0=F_{\overline{v_{iR}}}&=&M_{\Delta ij}v_{jR}+f_{lk}^i\psi_l\psi_k ,  \nonumber\\
0=F_{{v_{jR}}}&=&\overline{v_{iR}} M_{\Delta ij}+g_{lk}^j\overline{\psi}_k\overline{\psi}_l ,      \nonumber\\
0=F_{\overline{\psi}_k}&=&M_{\psi_{kl}}\psi_l+g_{lk}^jv_{jR}\overline{\psi}_l ,       \nonumber\\
0=F_{\psi_l}&=&\overline{\psi}_k M_{\psi_{kl}}+f_{lk}^i\overline{v_{iR}}\psi_k .      \label{equv4}
\end{eqnarray}
Explicitly, the second equation in (\ref{equv4}) gives
\be
\overline{v_{1R}}=-g_{lk}^2\overline{\psi}_k\overline{\psi}_l (M^{-1})_{21},  ~~
\overline{v_{2R}}=-g_{lk}^1\overline{\psi}_k\overline{\psi}_l (M^{-1})_{12},\nonumber
\ee
then, to generate a VEV of the  seesaw scale for $\overline{v_{1R}}$,
at least one VEV $\overline{\psi}$ is lower than $\Lambda_{GUT}$.
If there is only one $\overline{\psi}$, the second equation gives also a low $\overline{v_{2R}}$.
Then the D-flatness condition (\ref{Dflat}) suggests that the $U(1)_{B-L}$ breaks at a scale lower than
$\Lambda_{GUT}$, which violates gauge coupling unification.
Thus we must have at least two pairs
${\Psi}_{4,5}+\overline{\Psi}_{4,5}$ in the spinor representations whose labels are
 different from the matter superfields $\Psi_{1,2,3}$.
Furthermore,
we can diagonalize the coupling $g_{lk}^2$ so that only $g_{44}^2\neq 0$ without lost of generality.
Also, the coupling $f^1_{kl}$ must be zero so that $\Psi_{4,5}$  are different from  $\Psi_{1,2,3}$
which  couple with $\overline{\Delta}_1$ through the Yukawa couplings.

We have tried using many different forms of $M_{\Psi_{kl}}$'s
for all F-flatness conditions fulfilled.
We find the following successful model by introducing two SO(10) singlets $S, S^\prime$ and one more $A^\prime(\textbf{45})$
in addition to the original $A(\textbf{45})+E(\textbf{54})$.
Suppressing all the dimensionless couplings, the full superpotential for the GUT symmetry breaking is
\begin{eqnarray} \label{superp1}
W&=&\overline{\Delta}_1M_{\Delta 12}\Delta_2+\overline{\Delta}_2M_{\Delta 21}\Delta_1
+S\overline{\Psi}_4\Psi_5
+m_{54}\overline{\Psi}_5\Psi_4+A'\overline{\Psi}_5\Psi_5
+\Delta_1\overline{\Psi}_5^2+\overline{\Delta}_2\Psi_4^2 \nonumber \\
&+&\Delta_2\overline{\Psi}_4^2+S'A'A+SA'^2
+\frac{1}{2}m_AA^2+\frac{1}{2}m_EE^2+A^2E+E^3,
\end{eqnarray}
which is protected by an anomalous $U(1)_A$ under which the charges of the superfields
are listed in Table 1.
Here explicitly
$M_{\Delta ij}=m_{\Delta ij}-\lambda_{ij}\frac{1}{5} A_1-\lambda_{ij}\frac{\sqrt{3}}{5\sqrt{2}}A_2$.
\begin{table}[h]\scriptsize
\begin{center}
\begin{tabular}{|c|c|c|c|c|c|c|c|c|c|c|c|c||c|}
\hline
 &$\overline{\Delta}_1$ & $\Delta_1$ & $\overline{\Delta}_2$ & $\Delta_2$ & $\overline{\Psi}_4$ & $\Psi_4$ & $\overline{\Psi}_5$ & $\Psi_5$ & $A,E$ & $S$ & $S'$ & $A'$ & $\Psi_{1,2,3}$ \\
\hline
charge & $-6b-4c$ & $-2c$ & $2c$ & $6b+4c$ & $-3b-2c$ & $-c$ & $c$ & $b$ & 0 & $2b+2c$ & $b+c$ & $-b-c$ & $3b+2c$ \\
\hline
\end{tabular}
\caption{SO(10) multiplets and their $U(1)_A$ charges.} \label{Qnumbers}
\end{center}
\end{table}

In string models
the anomalous $U(1)_A$ symmetry is only anomalous in the effective theory below the string scale.
The D-term of such an $U(1)_A$ symmetry
gets a non-zero Fayet-Iliopoulos term $\xi$ related to the string scale as\cite{green1,green2,green3,green4}
\be
D_A =- \xi + \sum Q_i |\varphi_i|^2 ,
~~~~~
\xi= \frac{M_{st}^2}{192\pi^2} {\rm Tr}Q,
\ee
where the sum includes all scalar fields $\varphi_i$ present
in the theory with nonzero $U(1)_A$ charges $Q_i$.
Then the SO(10) singlet gets a non-zero VEV\cite{green11,green12}
\be
S \sim 10^{-1} M_{st}\sim 10M_G\label{SVEV}
\ee
from the anomalous D-term to preserve SUSY.
Numerically a variation of order one in $S$ is reasonable in (\ref{SVEV}).

Now the F-flatness conditions are
\begin{eqnarray}
0=F_{\overline{v_{1R}}}&=&M_{12}v_{2R},   \nonumber\\
0=F_{v_{2R}}&=&M_{21}\overline{v_{1R}}+\overline{\psi}_4^2,   \nonumber\\
0=F_{\overline{v_{2R}}}&=&M_{21}v_{1R}+\psi_4^2,   \nonumber\\
0=F_{v_{1R}}&=&M_{21}\overline{v_{2R}}+\overline{\psi}_5^2,   \nonumber\\
0=F_{\overline{\psi}_4}&=&S\psi_5+2\overline{\psi}_4v_{2R},   \nonumber\\
0=F_{\psi_4}&=&m_{54}\overline{\psi}_5+2\psi_4\overline{v_{2R}},   \nonumber\\
0=F_{\overline{\psi}_5}&=&m_{54}\psi_4-2A_1'\psi_5-\sqrt{6}A_2'\psi_5+2v_{1R}\overline{\psi}_5,   \nonumber\\
0=F_{\psi_5}&=&S\overline{\psi}_4-2A_1'\overline{\psi}_5-\sqrt{6}A_2'\overline{\psi}_5,   \nonumber\\
0=F_S&=&\overline{\psi}_4\psi_5+A_1'^2+A_2'^2,   \nonumber\\
0=F_{S'}&=&A_1'A_1+A_2'A_2,   \nonumber\\
0=F_{A'_1}&=&-2\overline{\psi}_5\psi_5+S'A_1+2SA'_1,   \nonumber\\
0=F_{A'_2}&=&-\sqrt{6}\overline{\psi}_5\psi_5+S'A_2+2SA'_2,   \nonumber\\
0=F_{A_1}&=&m_A A_1+\sqrt{\frac{3}{5}}A_1E+S'A_1'
-\frac{1}{5} \overline{v_{1R}}v_{2R}-\frac{1}{5} \overline{v_{2R}}v_{1R},\nonumber\\
0=F_{A_2}&=&m_A A_2-\frac{2}{\sqrt{15}}A_2E+S'A_2'
-\frac{\sqrt{3}}{5\sqrt{2}} \overline{v_{1R}}v_{2R}-\frac{\sqrt{3}}{5\sqrt{2}} \overline{v_{2R}}v_{1R},  \nonumber\\
0=F_E&=&m_EE+\frac{\sqrt{3}}{2\sqrt{5}}E^2+\frac{\sqrt{3}}{2\sqrt{5}}A_1^2-\frac{1}{\sqrt{15}}A_2^2.\label{Fterms}
\end{eqnarray}
Solving all these equations gives
one set of the solutions which require $v_2=0$ and $\psi_5=0$
and give the relations
\be
\overline{v_{1R}}=-\frac{\overline{\psi}_4^2}{M_{12}},
~~\overline{\psi}_4=\frac{2A_1'+\sqrt{6}A_2'}{S}\overline{\psi}_5,
~~A_{1,2}'=-\frac{S'}{S}A_{1,2}.\nonumber
\ee
Taking $S=10M_G$, the  other VEVs are naturally
\bea
\begin{array}{cccc}
\overline{v}_{1R}\sim 10^{-3}M_G, &\overline{v}_{2R}\sim M_G, &v_{1R}\sim M_G, &v_{2R}=0, \\
\overline{\psi}_4\sim 10^{-\frac{3}{2}}M_G, &\overline{\psi}_5\sim M_G, &\psi_4\sim M_G, &\psi_5=0,\\
E\sim M_G, &A_{1,2}\sim M_G, &A_{1,2}^\prime \sim 10^{-\frac{1}{2}}M_G, &S^\prime\sim 10^{\frac{1}{2}}M_G.\\
\end{array}
\eea
Now a VEV $\overline{v}_1\sim 10^{13}$GeV is generated as the seesaw scale,
a factor of 10 smaller than that got in \cite{lz}.
However, if in (\ref{SVEV}) $S$ is taken a smaller value, then
this seesaw scale can be $\sim 10^{14}$GeV now, comparable to that in \cite{lz}.

It can be also checked that the Goldstone mode for the $B-L$ breaking has components only from
$\overline{\Delta}_{1,2}, \Delta_1, \overline{\Psi}_{4,5}, \Psi_4$,
\be
\frac{2v_{1R}}{N}\Delta_1-\frac{2\overline{v}_{1R}}{N}\overline{\Delta}_1
-\frac{2\overline{v}_{2R}}{N}\overline{\Delta}_2
+\frac{\psi_4}{N}\Psi_4-\frac{\overline{\psi}_4}{N}\overline{\Psi}_4
-\frac{\overline{\psi}_5}{N}\overline{\Psi}_5,
\ee
by the F-flatness conditions.
This agrees with the observation made in Section 4
that those fields with null $B-L$ breaking VEVs cannot enter into the Goldstone mode.

In the full model, $H_1(\textbf{10})$ and $D_1(\textbf{120})$ are introduced to have the same $U(1)_A$ charges as $\overline{\Delta}_1$'s.
We introduce $H_2$ and $D_2$ with the same charges as ${\Delta}_2$'s.
We also impose a $Z_2$ symmetry, under which only the matter superfields $\Psi_{1,2,3}$ are odd,
to suppress unwanted couplings.
The full Higgs superpotential consistent with the $U(1)_A\times Z_2$ symmetry is
\bea
&&m_{H12}H_1H_2+m_{D12}D_1D_2+m_{\Delta 12}\overline{\Delta}_1{\Delta}_2+m_{\Delta 21}\overline{\Delta}_2 {\Delta}_1
+EH_1H_2+ED_1D_2\nonumber\\
&+&A\left( H_1H_2+H_1D_2+H_2D_1+D_1D_2
+D_1\overline{\Delta}_2+D_1{\Delta}_2+D_2\overline{\Delta}_1+D_2{\Delta}_1
+\overline{\Delta}_1{\Delta}_2+\overline{\Delta}_2{\Delta}_1\right)
\nonumber\\
&+&S\overline{\Psi}_4\Psi_5
+m_{54}\overline{\Psi}_5\Psi_4+A'\overline{\Psi}_5\Psi_5
+\Delta_1\overline{\Psi}_5^2+\overline{\Delta}_2\Psi_4^2
+(H_2+D_2+\Delta_2)\overline{\Psi}_4^2\nonumber \\
&+&S'A'A+SA'^2
+\frac{1}{2}m_AA^2+\frac{1}{2}m_EE^2+A^2E+E^3.
\eea

To study proton decay we need to know
the full color-triplet mass matrix.
The columns are ordered as {$H_1$, $D_1$ ,$D_1'$, $\overline{\Delta}_1$, $\overline{\Delta}_1'$, $\Delta_1$; $H_2$, $D_2$, $D_2'$, $\Delta_2$,
$\overline{\Delta}_2$, $\overline{\Delta}_2'$; $\overline{\psi}_4$, $\overline{\psi}_5$},
and the rows are ordered as the conjugations,
\begin{equation} \label{TRM}
M_T=\left(
\begin{array}{ccc}
0_{(6\times 6)} & B_{12(6\times 6)} & B_{13(6\times 2)}\\
B_{21(6\times 6)} & 0_{(6\times 6)} & B_{23(6\times 2)}\\
0_{(2\times 6)} & B_{32(2\times 6)} & B_{33(2\times 2)}
\end{array}
\right),
\end{equation}
where
\begin{equation}\label{B12}
B_{12}=\left(
\begin{array}{cccccc}
m_H-\frac{E}{\sqrt{15}}+i\frac{A_2}{\sqrt{6}} & -\frac{i\sqrt{2}}{3}A_2 & -\frac{i}{\sqrt{3}}A_1 & 0 & 0 & 0\\
\frac{i\sqrt{2}}{3}A_2 & -\frac{E}{\sqrt{15}} & 0 & \frac{iA_2}{\sqrt{30}} & 0 & 0\\
\frac{iA_1}{\sqrt{3}} & \frac{2E}{3\sqrt{15}} & -\frac{iA_1}{2\sqrt{5}} & 0 & 0\\
0 & -\frac{iA_2}{\sqrt{30}} & \frac{iA_1}{2\sqrt{5}} & m_{12}-\frac{iA_2}{5\sqrt{6}} & 0 & 0\\
0 & 0 & 0 & 0 & m_{21}-\frac{iA_2}{5\sqrt{6}} & 0\\
0 & 0 & 0 & 0 & 0 & m_{21}-\frac{iA_2}{5\sqrt{6}}
\end{array}
\right),\nonumber
\end{equation}
\begin{equation}\label{B13}
B_{13}=\left(
\begin{array}{cc}
0 & 0\\
0 & 0\\
0 & 0\\
0 & 0\\
0 & -i\sqrt{3}\overline{\psi}_5\\
0 & \sqrt{6}\overline{\psi}_5
\end{array}
\right),\nonumber
\end{equation}
\begin{equation}\label{B21}
B_{21}=\left(
\begin{array}{cccccc}
m_H-\frac{iA_2}{\sqrt{6}}+\frac{E}{\sqrt{15}} & -\frac{i\sqrt{2}}{3}A_2 & -\frac{i}{\sqrt{3}}A_1 & 0 & 0 & 0\\
\frac{i\sqrt{2}}{3}A_2 & m_D-\frac{E}{\sqrt{15}}-\frac{iA_2}{3\sqrt{6}} & 0 & -\frac{iA_2}{\sqrt{30}} & -\frac{iA_1}{\sqrt{10}} & -\frac{iA_2}{\sqrt{30}}\\
\frac{iA_1}{\sqrt{3}} & 0 & m_D+\frac{2E}{3\sqrt{15}}-\frac{iA_2}{3\sqrt{6}} & -\frac{iA_1}{2\sqrt{5}} & -\frac{iA_2}{\sqrt{15}} & -\frac{iA_1}{2\sqrt{5}}\\
0 & -\frac{iA_2}{\sqrt{30}} & \frac{iA_1}{2\sqrt{5}} & 0 & 0 & m_{21}-\frac{iA_2}{5\sqrt{6}}\\
0 & \frac{iA_2}{\sqrt{30}} & \frac{iA_1}{2\sqrt{5}} & m_{12}-\frac{iA_2}{5\sqrt{6}} & 0 & 0\\
0 & \frac{iA_1}{\sqrt{10}} & \frac{iA_2}{\sqrt{15}} & 0 & m_{12}-\frac{iA_2}{5\sqrt{6}} & 0
\end{array}
\right),\nonumber
\end{equation}
\begin{equation}\label{B23}
B_{23}=\left(
\begin{array}{cc}
-\overline{\psi}_4 & 0\\
\sqrt{2}\overline{\psi}_4 & 0\\
-i\sqrt{2}\overline{\psi}_4 & 0\\
0 & 0\\
-i\sqrt{3}\overline{\psi}_4 & 0\\
\sqrt{6}\overline{\psi}_4 & 0
\end{array}
\right),\nonumber
\end{equation}
\begin{equation}\label{B32}
B_{32}=\left(
\begin{array}{cccccc}
0 & 0 & 0 & 0 & i\sqrt{3}\psi_4 & \sqrt{6}\psi_4\\
0 & 0 & 0 & 0 & 0 & 0
\end{array}
\right),\nonumber
\end{equation}
and
\begin{equation}\label{B33}
B_{33}=\left(
\begin{array}{cc}
0 & m_{54}-\sqrt{6}A_2-2A_1\\
S & -\sqrt{6}A_2'-2A_1'
\end{array}
\right).\nonumber
\end{equation}
The proton decay rates depend only on the effective triplet mass matrix corresponding to
the superfields which appear in the Yukawa couplings (\ref{yuka}).
Integrating out those superfields which are absent in  (\ref{yuka}),
we find that that all entries in this effective triplet mass matrix are infinities
so that there is no color higgsino mediated proton decay and proton decays
are all from gauge mediation in the present model.
This is quite different from the model
in \cite{dlz2,lz,cz2} where although the dimension-five operators are safe from the data,
but in general they dominate over the mechanism of gauge mediation\cite{czy}.
However, since we have not dealt with the DTS problem in the present model,
this conclusion need to be taken carefully in future studies.

\section{Summary}
In this paper, we have proposed an alternative  model
to generate the seesaw scale.
Proton decays through dimension-five operators are absent which are
very different from the models studied before.
However, further work needs to be done to solve the DTS problem in the present model.

\newpage
\appendix
\section*{Appendix}
\section{The Goldstone mode for $U(1)$ symmetry breaking}

Only those superfields $X$'s which contain the SM singlets may contribute to the GUT
and especially to the
$U(1)_{B-L}$  symmetry breaking.
For a general consideration, the superpotential with $m$ different $X$'s is
\be
W=\sum_{n_1\cdots n_m} f_{n_1\cdots n_m} X_1^{n_1}\cdots X_m^{n_m},
\ee
with the conservation of the  $U(1)$ charges
\be
0=f_{n_1\cdots n_m}\sum_{a=1}^m n_a q_a.\label{Scharge}
\ee
Hereon $X_a$   represents  the VEV and its
 F-flatness condition is
\be
0=F_{X_a}=\sum_{n_1\cdots n_m} \frac{n_a}{X_a} f_{n_1\cdots n_m} X_1^{n_1}\cdots X_m^{n_m},\label{SFTE}
\ee
and the  mass matrix  elements of the SM singlets are
\be
M_{ab}=\frac{\partial^2 W}{\partial X_a \partial X_b}=
\sum_{n_1\cdots n_m}\frac{n_a n_b}{X_a X_b}(1 -\frac{\delta _{a b}}{n_b})
f_{n_1\cdots n_m} X_1^{n_1}\cdots X_m^{n_m}.
\ee
Acting the mass matrix on the column vector
\be
\hat{G}=(q_1 X_1, \cdots, q_b X_b, \cdots, q_m X_m)^{\rm T}\label{Gmode}
\ee
gives a column vector whose  component is
\bea
\sum_{b=1}^m M_{ab} G_b&=&\sum_{b=1}^m\sum_{n_1\cdots n_m} \frac{n_a n_b q_b}{X_a}(1 -\frac{\delta _{a b}}{n_b})
f_{n_1\cdots n_m} X_1^{n_1}\cdots X_m^{n_m} \\\nonumber
&=& \sum_{n_1\cdots n_m} \frac{n_a}{X_a}\left[f_{n_1\cdots n_m}\sum_{b=1}^m n_b q_b\right]  X_1^{n_1}\cdots X_m^{n_m} -q_a\left[\sum_{n_1\cdots n_m}\frac{n_a}{X_a}f_{n_1\cdots n_m} X_1^{n_1}\cdots X_m^{n_m}\right]\\\nonumber
&=&0
\eea
following (\ref{Scharge}) and (\ref{SFTE}).
Then $\hat{G}$ in (\ref{Gmode}) is the Goldstone mode of the $U(1)$ symmetry breaking.


\begin{thebibliography}{99}
\bibitem{so10a}T.~E.~Clark, T.~K.~Kuo and N.~Nakagawa,
\textit{An  SO(10) supersymmetric grand unified theory},
\textit{Phys.~Lett.}~\textbf{B~115}~(1982)~26.

\bibitem{so10b}
C.~S.~Aulakh and R.~N.~Mohapatra,
\textit{Implications of supersymmetric SO(10) grand unification},
\textit{Phys.~Rev.}~\textbf{D~28}~(1983)~217.

\bibitem{unif1} M. B. Einhorn and D. R. T. Jones, \textit{The Weak Mixing Angle and Unifi-
cation Mass in Supersymmetric} SU(5), \textit{Nucl. Phys.} \textbf{B 196} (1982) 475.

\bibitem{unif2}
W. J. Marciano and G. Senjanovi$\acute{\textrm{c}}$, \textit{Predictions of Supersymmetric Grand Unified Theories}, \textit{Phys. Rev.} \textbf{D 25} (1982) 3092.

\bibitem{MSSM1} U. Amaldi, W. de Boer and H. Furstenau, \textit{Comparison of grand unified theories
with electroweak and strong coupling constants measured at LEP}, \textit{Phys. Lett.} \textbf{B 260} (1991) 447.

\bibitem{MSSM2} John Ellis, S. Kelley and D. V. Nanopoulos, \textit{Probing the desert using gauge coupling unification}, \textit{Phys. Lett.} \textbf{B 260} (1991) 131.

\bibitem{MSSM3} P. Langacker and M. Luo, \textit{Implications of precision electroweak experiments for $m_t$, $\rho_0$, $sin^2\theta_{\omega}$, and grand unification}, \textit{Phys. Rev.} \textbf{D 44} (1991) 817.

\bibitem{MSSM4} C. Giunti, C. W. Kim, and U. W. Lee, \textit{Running coupling constants and grand unification models}, \textit{Mod. Phys. Lett.} \textbf{A 06} (1991) 17.


\bibitem{seesawi1}
P.~Minkowski, $\mu\to e\gamma$ \textit{at a rate of one out of 109 muon decays}?, \textit{Phys.~Lett.}~\textbf{B~67}~(1977)~421.

\bibitem{seesawi2}
T.~Yanagida, in \textit{workshop on unified theories}, KEK Report 79-18~(1979)~95.

\bibitem{seesawi3}
M. Gell-Mann, P. Ramond, and R. Slansky, in  \textit{Supergravity}, P. van Nieuwenhuizen and D. Z. Freedman (eds.), North Holland Publ. Co, (1979) p. 315.

\bibitem{seesawi4}
S. L. Glashow, in \textit{1979 Cargese Summer Institute on Quarks and Leptons}, Plenum, New York, (1980) p. 687.

\bibitem{seesawi5}
R. N. Mohapatra and G. Senjanovi$\acute{\textrm{c}}$, \textit{Neutrino mass and spontaneous parity nonconservation}, \textit{Phys. Rev. Lett.} \textbf{44} (1980) 912.

\bibitem{seesawii1}
G. Lazarides, Q. Shafi, and C. Wetterich, \textit{Proton lifetime and fermion masses in an} SO(10) \textit{model}, \textit{Nucl. Phys.} \textbf{B 181} (1981) 287.

\bibitem{seesawii2}
R. N. Mohapatra and G. Senjanovi$\acute{\textrm{c}}$, \textit{Neutrino masses and mixings in gauge models with spontaneous parity violation}, \textit{Phys. Rev.} \textbf{D 23} (1981) 165.

\bibitem{seesawii3}
J. Schechter and J. W. F. Valle, \textit{Neutrino masses in} SU(2)$\otimes$U(1) \textit{theories}, \textit{Phys. Rev.} \textbf{D 22} (1980) 2227.

\bibitem{seesawii4}
E. Ma and U. Sarkar, \textit{Neutrino masses and leptogenesis with heavy Higgs triplets}, \textit{Phys. Rev. Lett.} \textbf{80} (1998) 5716.

\bibitem{sato}
J. Sato, \textit{A SUSY SO(10) GUT with an intermediate scale},
\textit{Phys. Rev.} \textbf{D 53} (1996) 3884.

\bibitem{aulakh4554}
C. S. Aulakh, B. Bajc, A. Melfo, A. Rasin, (North Carolina U.), snf G. Senjanovi$\acute{\textrm{c}}$,
\textit{SO(10) theory of R-parity and neutrino mass },
\textit{Nucl. Phys.} \textbf{B 597} (2001) 89.

\bibitem{msso10a}
C. S. Aulakh, B. Bajc, A. Melfo, G. Senjanovi$\acute{\textrm{c}}$, and F. Vissani, \textit{The minimal supersymmetric grand unified theory}, \textit{Phys. Lett.} \textbf{B 588} (2004) 196.

\bibitem{msso10b}
T. Fukuyama, T. Kikuchi, A. Ilakovac, S. Meljanac, and N. Okada, \textit{Detailed analysis of proton decay rate in the minimal supersymmetric} SO(10) \textit{model}, \textit{JHEP} 0409 (2004) 052.

\bibitem{bajc}
B. Bajc, A. Melfo, G. Senjanovi$\acute{\textrm{c}}$, and F. Vissani, \textit{Minimal supersymmetric grand unified theory: Symmetry breaking and the particle spectrum}, \textit{Phys. Rev.} \textbf{D 70} (2004) 035007.

\bibitem{msso10d}
T. Fukuyama, A. Ilakovac,  T. Kikuchi, S. Meljanac, and N. Okada,  SO(10) \textit{Group theory for the unified model building},  \textit{J. Math. Phys.} \textbf{46} (2005)  033505.

\bibitem{bert}
S. Bertolini, T. Schwetz and M. Malinsky,
\textit{Fermion masses and mixings in SO(10) models and the neutrino challenge to SUSY GUTs},
\textit{Phys. Rev.} \textbf{D~73} (2006) 115012.

\bibitem{malinsky}
M. Malinsk$\acute{\textrm{y}}$,
\textit{Higgs sector of the next-to-minimal renormalizable SUSY SO(10)},

\bibitem{aulakh120}
C. S. Aulakh and S. K. Garg,
\textit{The New Minimal Supersymmetric GUT : Spectra, RG analysis and Fermion Fits},
\textit{Nucl.Phys.} \textbf{ B~857} (2012) 101.

\bibitem{su5pd1}
N. Sakai and T. Yanagida, \textit{Proton decay in a class of supersymmetric grand unified models}, \textit{Nucl. Phys.} \textbf{B 197} (1982) 533.

\bibitem{su5pd2}
S. Weinberg, \textit{Supersymmetry at ordinary energies. Masses and conservation laws}, \textit{Phys. Rev.} \textbf{D 26} (1982) 287.

\bibitem{su5pd3}
P. Nath, A. H. Chamseddine and R. L. Arnowitt,
\textit{Nucleon Decay in Supergravity Unified Theories},
\textit{Phys. Rev.} \textbf{D 32} (1985) 2348.

\bibitem{su5pd4}
J.~Hisano, H.~Murayama and T.~Yanagida,
\textit{Nucleon decay in the minimal supersymmetric SU(5) grand unification,}
\textit{Nucl. Phys.} \textbf{B 402} (1993) 46.


\bibitem{cz2}
Y.-K. Chen and D.-X. Zhang,
\textit{A renormalizable supersymmetric SO(10) model},
\textit{JHEP} \textbf{1507} (2015) 005.

\bibitem{dlz2}
L. Du, X. Li and D.-X. Zhang,
\textit{Connection between proton decay suppression and seesaw mechanism in supersymmetric SO(10) models},
\textit{JHEP} \textbf{1410} (2014) 036.

\bibitem{dw1}S. Dimopoulos and F. Wilczek,
\textit{Incomplete Multiplets in Supersymmetric Unified Models},
 NSF-ITP-82-07.

\bibitem{dw2}Mark Srednicki,
\textit{Supersymmetric Grand Unified Theories and the Early Universe},
\textit{Nucl.Phys.} \textbf{ B~202} (1982) 327.

\bibitem{dw3}
K. S. Babu and S. M. Barr,
\textit{Natural suppression of Higgsino mediated proton decay in supersymmetric SO(10)},
 \textit{Phys.~Rev.}~\textbf{D~48} (1993) 5354.

\bibitem{lee}
 D.-G. Lee and R.N. Mohapatra,
 \textit{Natural doublet - triplet splitting in supersymmetric SO(10) models},
\textit{Phys.~Lett.}~\textbf{B~324} (1994) 376.
\textit{JHEP}~\textbf{1006} (2010) 084.

\bibitem{cz}
Y.-K. Chen and D.-X. Zhang,
\textit{A renormalizable supersymmetric SO(10) model with natural doublet-triplet splitting},
\textit{JHEP} \textbf{1501} (2015) 025.


\bibitem{czy}
Z.-Y. Chen and D.-X. Zhang, \textit{Examining A Renormalizable Supersymmetric SO(10) Model},
arXiv:1611.07760.

\bibitem{lz}
X. Li and D.-X. Zhang, \textit{Proton decay suppression in a supersymmetric} SO(10) \textit{model}, \textit{JHEP} 02(2015) 130.

\bibitem{green1}
M. B. Green and J. H. Schwarz,
\textit{Anomaly cancellations in supersymmetric $D=10$ gauge theory and superstring theory},
\textit{Phys. Lett.} \textbf{B 149} (1984) 117.

\bibitem{green2}
M. Dine, N. Seiberg  and E. Witten,
\textit{Fayet-Iliopoulos terms in string theory}, \textit{Nucl. Phys.}
\textbf{B 289} (1987) 589.

\bibitem{green3}
J. J. Atick, L. J. Dixon  and A. Sen,
\textit{String calculation of fayet-iliopoulos D-terms in arbitrary supersymmetric compactifications},
\textit{Nucl. Phys.} \textbf{B 292} (1987) 109.

\bibitem{green4}
M. Dine, I. Ichinose  and N. Seiberg,
\textit{F terms and D terms in string theory},
\textit{Nucl. Phys.} \textbf{B 293} (1987) 253.


 \bibitem{green11}
 Z. Berezhiani and Z. Tavartkiladze,
 \textit{More missing VEV mechanism in supersymmetric SO(10) model},
 \textit{Phys.~Lett.}~\textbf{B~409} (1997) 220.


\bibitem{green12}
K. S. Barr, Jogesh C. Pati and Zurab Tavartkiladze, \textit{Constraining Proton Lifetime in SO(10) with Stabilized Doublet-Triplet Splitting}, \textit{JHEP} 06(2010) 084.

\bibitem{aulakh2005}
C. S. Aulakh and A. Girdhar, \textit{SO(10) MSGUT : spectra, couplings and threshold effects}, \textit{Nucl.Phys.} \textbf{B 711} (2005) 275.

\bibitem{babu2012}
K. S. Babu and R. N. Mohapatra, \textit{B-L Violating Proton Decay Modes and New Baryogenesis Scenario in SO(10)}, arXiv:1207.5771; \textit{B-L Violating Nucleon Decay and GUT Scale Baryogenesis in SO(10)}, arXiv:1203.5544.

\end{thebibliography}
\end{document}